\documentstyle[eqsecnum,aps]{revtex}
\begin{document}
\draft
\title{Pseudoscalar Meson Mixing in Effective Field Theory}
\author{ E.Gedalin\thanks{gedal@bgumail.bgu.ac.il},
 A.Moalem\thanks{moalem@bgumail.bgu.ac.il}
 and L.Razdolskaya\thanks{ljuba@bgumail.bgu.ac.il}}
\address
{ Department of Physics, Ben Gurion University, 84105,
Beer Sheva, Israel}
\maketitle
\begin{abstract}
We show that for any effective field theory of colorless meson fields, the
mixing schemes of particle states and decay constants are not only
related but also determined exclusively by the kinetic and mass Lagrangian
densities. In the general case, these are bilinear in terms of the
intrinsic fields and involve non-diagonal kinetic and mass matrices.
By applying three consecutive steps this 
Lagrangian can be reduced into the standard quadratic form 
in terms of the physical fields. These steps are : (i) the diagonalization
of the kinetic matrix, (ii) rescaling of the fields, and (iii) the 
diagonalization of the  mass matrix. In case, where 
the dimensions of the non-diagonal kinetic and mass sub-matrices  
are respectively, $k\times k$ and $n\times n$, this procedure leads to 
mixing schemes which involve $[k(k-1)/2] + [n(n-1)/2]$ angles and $k$ 
field rescaling parameters. This observation holds true irrespective with
the type of particle interactions presumed. The commonly used mixing
schemes, correspond to a proper choice of the kinetic and mass 
matrices, and are derived as special cases. In particular, 
$\eta$-$\eta '$ mixing, requires one angle, if and only if, the kinetic
term with the intrinsic fields has a quadratic form.
\end{abstract}
\ \\

Mixing schemes of the pseudoscalar $\eta (547.3 MeV)-\eta' (957 MeV)$
mesons has attracted considerable interest in recent
years\cite{gilman87}-\cite{gedalin00}. The
traditional approach based on the flavor $SU(3)$ quark model\cite{pdg00} 
involves a single mixing angle $\theta_P$, an octet ($F_8$) and a 
singlet ($F_0$) radiative decay constants.  The physical
states $\eta ~\mbox{and}~\eta '$ are taken to be linear
combinations of the octet $\eta_8$ and  singlet $\eta_0$ states, i.e.,
\begin{equation}
 \left(\begin{array}{c}
		\eta  \\
		\eta '
	\end{array}\right) = \Omega(\theta_P)\left(\begin{array}{c}
		\eta_8 \\
		\eta_0
	\end{array}\right)~,
\label{oneang}
\end{equation}
where $\Omega(\theta_P)$ stands for the unitary matrix,
\begin{equation}
\Omega(\theta_P)=\left(
	\begin{array}{cc}
	\cos\theta_P & -\sin\theta_P  \\
	\sin\theta_P & \cos\theta_P
	\end{array}¥
	\right)~.
	\label{omega}
\end{equation}
The corresponding physical decay constants $F_\eta$ and $F_{\eta '}$ 
are presumably related to the octet and singlet decay constants through
the same unitary transformation. The values of the mixing
angle and decay constants are determined from phenomenological ( model
dependent) data analyses of various processes 
\cite{gilman87,bramon90,ball96,bramon99,benayoun98,benayoun981,veno98},
but with a wide range of uncertainty.  It is $ab~initio$ not clear why
the state and decay constant mixing schemes should be identical or
even similar.  Indeed, several authors
\cite{schechter93,kisselev93,leut97,kaiser98,ali98,feld981} suggested
a two mixing angle scheme for the decay constants, e.g.,
\begin{equation}
	\left(
	\begin{array}{cc}
		F^8_\eta & F^0_\eta  \\
		F^8_{\eta '} & F^0_{\eta '}
	\end{array}¥
	\right) = \bar{\Theta}(\theta_{\eta}, \theta_{\eta'})\left(
	\begin{array}{cc}
		F_8 & 0  \\
		0 & F_0
	\end{array}¥
	\right); \qquad \bar{\Theta}(\theta_{\eta}, \theta_{\eta'}) = \left(
	\begin{array}{cc}
	\cos\theta_{\eta} & -\sin\theta_{\eta'} \\
	\sin\theta_{\eta} & \cos_{\eta'}
	\end{array}
	\right)~,
	\label{twoang}
\end{equation}¥
while the particle states follow either the same transformation
\cite{escrib99} or a single angle mixing 
pattern\cite{leut97,kaiser98,feld981} as in Eqn.\ref{oneang}. 
To explain data, the two mixing angles $\theta_{\eta}$ and
$\theta_{\eta'}$ turned out to be considerably
different\cite{leut97,kaiser98,escrib99}.  Recently, Feldmann et
al.\cite{feld98,feld98a,feld98b} proposed that only in the quark
flavor basis (QFB), $q\bar{q} = (u\bar{u} +d\bar{d})/\sqrt{2}$ and
$s\bar{s}$, the decay constants and particle states follow the same
mixing pattern, namely,
\begin{equation}
		\left(
	\begin{array}{cc}
		F^q_\eta & F^s_\eta  \\
		F^q_{\eta '} & F^s_{\eta '}
	\end{array}¥
	\right) = \Omega(\phi)\left(
	\begin{array}{cc}
		F_q & 0  \\
		0 & F_s
	\end{array}¥
	\right), 
	\label{ffeld}
\end{equation}¥
and,
\begin{equation}
	 \left(\begin{array}{c}
		\eta  \\
		\eta '
	\end{array}\right) = \Omega(\phi)\left(\begin{array}{c}
		\eta_ q\\
		\eta_s
	\end{array}\right)~.
	\label{mfeld}
\end{equation}
Clearly, the advantage of such a  scheme is 
that only one mixing angle $\phi$ is required. It is to be stressed that 
by assumption this simplification is restricted to the case of two
orthogonal QFB states only. Usually, in the 
octet-singlet basis, 
however, there is a need for one mixing angle
$\theta_P = \phi - \theta_{ideal},~~\theta_{ideal} = \arctan\sqrt{2}$ 
for the particle states and two mixing angles,
 Eqn. \ref{twoang}, for the decay constants. It is the purpose of 
the present note to show that for any effective 
field theory (EFT) irrespective with the field interactions presumed,
the mixing schemes for the states and decay constants
not only are related but also determined exclusively by
the structure of the kinetic and mass Lagrangian densities.

We recall that the QCD Lagrangian exhibits a $SU(3)_L \otimes SU(3)_R$
symmetry which breaks down spontaneously to $SU(3)_V$, giving rise to
an octet of light Goldstone pseudoscalar mesons. The axial $U(1)$ symmetry
of the QCD Lagrangian is broken by the anomaly, giving rise to a ninth
Goldstone pseudoscalar singlet meson. The corresponding octet and
singlet state mix because of $SU(3)$ flavor symmetry breaking, reflecting
a nontrivial nature of the QCD vacuum. Therefore, any EFT
of colorless fields which allows for a consistent treatment 
of hadronic physics must also describe the mixing and mass spectrum 
of the physical mesons. For such theories, the most general form of the
Goldstone meson kinetic and mass Lagrangian terms is, 
\begin{equation}
	L_{km} = 
	\frac{1}{2}(\partial_\mu \Phi)K(\partial^\mu \Phi) + 
	\frac{1}{2}\Phi M^2\Phi~,
	\label{lagkm}
\end{equation}
where $\Phi$ stands for the intrinsic field matrix, $K~\mbox{and}~M^2$ are 
the kinetic and mass matrices. Usually, $K~\mbox{and}~M^2$
are non-diagonal and  
the Lagrangian $L_{km}$  is bilinear rather than 
quadratic as invoked from the Klein-Gordon equation for the physical
fields. 
In what follows we show that this expression can always be reduced
into the standard quadratic form by applying three 
consecutive steps which transform the intrinsic fields
into the physical fields, and in turn determine the mixing schemes for
both the states and decay constants, uniquely.

As indicated already, the two matrices $K$ and
$M^2$ are, in general, non-diagonal. We first  
diagonalize $K$ using the unitary transformation, 
\begin{equation}
\Phi = \Upsilon \Phi'~.
\end{equation} 
Following this step the Lagrangian becomes,
\begin{equation}
	L_{km} = \frac{1}{2}(\partial_\mu \Phi')\hat{K}¥(\partial^\mu
\Phi') + 
		\frac{1}{2}\Phi'\Upsilon^{-1}{M}^2\Upsilon \Phi' ~,
	\label{lagkm1}
\end{equation}
with $\hat{K} = diag (\kappa_1, \kappa_2, \ldots)$. 
The eigenvalues
$\kappa_i (i=1,2,...)$, however, are not necessarily equal 1, and we
rescale the fields in order to restore the standard normalization 
of the kinetic term, i.e.,
\begin{equation}
	\Phi' = R\Phi ''~,
\end{equation}¥ 
where $R = diag(1/\sqrt{\kappa_1}, 1/\sqrt{\kappa_2},\ldots)$. 
With this second step,  
the kinetic term in terms of the renormalized fields $\Phi ''$ acquires 
the standard quadratic form and the mass matrix is,
\begin{equation}   
         \tilde{M}^2 = R\Upsilon^{-1}{M}^2\Upsilon R~.
        \label{}
\end{equation}
As a last step we diagonalize the matrix $\tilde{M}^2$ via another unitary
transformation of the fields $\Phi ''$ into the physical fields 
$\Phi_{ph}$,
\begin{equation}
       \Phi '' = \Omega \Phi_{ph}~.
\end{equation}
With this last step the transformed mass matrix becomes 
$M^2_{ph} = diag(m^2_1, m^2_2,...)$, where the eigenvalues $m_1, m_2,
\ldots$ are to be identified with the physical meson masses. 
Altogether the three steps transform the
intrinsic fields $\Phi$ into the  physical fields $\Phi_{ph}$,
\begin{equation}
        \Phi =  \Theta \Phi_{ph}~;~~~\Theta = \Upsilon R \Omega~. 
        \label{thetatr1}   
\end{equation}

We now turn to 
show that the meson decay constants $F_{ph}$  transform in a similar way. 
We start from the usual definition of the decay constants,
\begin{equation}
< 0| J^i_{\mu 5} | \Phi^m_{ph} > \equiv i F^{im}_{ph} q_{\mu}~,
\end{equation}
where $J^i_{\mu 5}$ stands for the axial vector currents, and the indices
$i~\mbox{and}~m$ label the intrinsic and physical fields.
From the Lagrangian Eqn. \ref{lagkm} and by substituting the
transformation
\ref{thetatr1} we may write,
\begin{equation}
     J_{\mu 5} =  F K \partial_\mu \Phi = F K \Theta \partial_\mu
\Phi_{ph}~,
        \label{axcur}
\end{equation}
where $F=diag(F_1, F_2, \ldots)$ is the
intrinsic decay constant matrix. Thus, the matrix of the physical decay
constants is,
\begin{equation}
	 F_{ph} = F K \Theta = \tilde{F} \Theta~,
	\label{¥}
\end{equation}
where $\tilde{F} = F K $ is the matrix of renormalized intrinsic decay
constants.

This complete our argument that the transformation $\Theta$ determines the
mixing schemes for the particle states and the decay constants.
Obviously the parameterization of this transformation  
depends on the dimensions of non-diagonal sub-matrices of $K~~\mbox{and}
~~M^2$.
Let the corresponding dimensions of these sub-matrices be
 $k\times k~\mbox{and}~n\times n$, respectively. We may parameterize 
the orthogonal transformation $\Upsilon$ using $k(k-1)/2$ real 
independent mixing angles. Likewise, the $\Omega$  would be described by 
$n(n-1)/2$ independent mixing angles. We may then conclude that the mixing
schemes involve at most, 
$[k(k-1)/2]+[n(n-1)/2]$ mixing angles and $k$ rescaling 
parameters, the value of which are determined by the matrix elements of
$K$ and $\tilde{M}^2$. 
In the special case where the kinetic term is non diagonal, but gets
diagonalized simultaneously with the mass matrix, the matrices commute
(i.e. $k=n$) and the number of mixing angles reduces to $k(k-1)/2$ only.
As an example, for the commonly considered
case of two intrinsic states $\eta_8$ and $\eta_0$, where 
$K$ is diagonal ($k=1$) and  $M^2$ is a $2\times 2$ matrix, only  
one mixing angle is required. Should we include a third component, like a
gluonium ($\eta_g$),  $K$ and $M^2$ are $3\times 3$ matrices. The most
commonly considered case is that corresponding to a diagonal kinetic
and a non diagonal mass matrices. Then the transformation $\Theta$
involves 
{\it a priori\/} 3 mixing angles.
With one of the mixing angles taken to be zero the gluonium component 
of one physical state vanishes. This leads to a 
two angle mixing scheme for the case of three particles.
By taking a second angle to be zero, one decouples the gluonium from 
the $\eta_8-\eta_0$ doublet, what yields again a one mixing angle pattern.
Clearly, however, for the most general case of  $\eta_8-\eta_0$ mixing
both $K$ and $M^2$ are  $2\times 2$ non diagonal matrices, and
therefore, the transformation $\Theta$ involves two mixing angles as 
well as two rescaling parameters.

Let us consider this last example in more details. The contributions of the 
$\eta_0$ and $\eta_8$ to the Lagrangian Eqn. \ref{lagkm} sum to,
       \begin{equation}
       	L_{km}^{08} = \frac{1}{2}
       	(
       	\begin{array}{cc}
       	\partial_\mu\eta_8	& \partial_\mu\eta_0
       	\end{array}¥
       	) K_{8,0} \left(
       	\begin{array}{c}
       		\partial^\mu\eta_8  \\
       		\partial^\mu\eta_0
       	\end{array}
       	\right) + \frac{1}{2}
       	(
       	\begin{array}{cc}
       \eta_8	&\eta_0
       	\end{array}¥
       	) {\cal M}^2(8,0) \left(
       	\begin{array}{c}
       		\eta_8  \\
       		\eta_0
       	\end{array}¥
       	\right)~,
       	\label{kinlag}
       \end{equation}¥  
with,
\begin{equation}
	K_{8,0}= \left(
	\begin{array}{cc}
		\kappa_{88} & \kappa_{08}  \\
		\kappa_{08} & \kappa_{00}
	\end{array}¥
	\right)~,\qquad {\cal M}^2(8,0)= \left(
	\begin{array}{cc}
		m^2_{88} & m^2_{08}  \\
		m^2_{08} & m^2_{00}
	\end{array}¥
	\right)~.
	\end{equation}
 
Following the procedure outlined above we first diagonalize the matrix
$K_{8,0}$ using the unitary transformation,
         \begin{equation}
         \left(\begin{array}{c}
                        \eta_8  \\
                        \eta_0
                \end{array}\right) 
                = \Upsilon\left(\begin{array}{c}
                        \bar{\eta}_8  \\
                        \bar{\eta}_0
                \end{array}\right)~,\qquad \Upsilon =  \left(\begin{array}{cc}
                        \cos\lambda & \sin\lambda  \\
                        -\sin\lambda & \cos\lambda
                \end{array}\right)~,
         \end{equation}  
which leads to,
\begin{equation}
    L_{km}^{08}  =
                \frac{1}{2}\kappa_8(\partial_\mu\bar{\eta}_8)^2 + 
                \frac{1}{2}\kappa_0(\partial_\mu\bar{\eta}_0)^2  + 
                 \frac{1}{2}(\bar{\eta}_8,\bar{\eta}_0)
                \Upsilon^{-1}{\cal M}^2\Upsilon
                \left(\begin{array}{c}
                        \bar{\eta}_8  \\
                        \bar{\eta}_0
                \end{array}\right)~.
         \end{equation}
Here $\kappa_0,~\mbox{and}~\kappa_8$ are eigenvalues of the matrix
$K_{8,0}$.
Next we  rescale the pseudoscalar fields according to,
         \begin{equation}
         \left(\begin{array}{c}
                       \bar{\eta}_8  \\
                        \bar{\eta}_0
                \end{array}\right)
                = R \left(\begin{array}{c}
                        \hat{\eta}_8  \\
                        \hat{\eta}_0
                \end{array}\right)~,\qquad
R = diag(z,~f) \equiv diag (1/\sqrt{\kappa_8}, 1/\sqrt{\kappa_0})~.
\end{equation}
Following these two 
transformations, the  kinetic term acquires the standard quadratic form,  
$(1/2)[ (\partial_\mu \hat{\eta}_8)^2 +
(\partial_\mu\hat{\eta}_0)^2]$ and the mass matrix becomes,
\begin{equation}
    \hat{{\cal M}}^2(8,0) = R\Upsilon^{-1}{\cal M}^2(8.0)\Upsilon R =
    \left(
    \begin{array}{cc}
        \hat{m}^2_{88} &  \hat{m}^2_{80} \\
        \hat{m}^2_{80} &  \hat{m}^2_{00}
    \end{array}
     \right).
    \label{}
\end{equation}
We may now 
diagonalize this matrix by another unitary transformation,
\begin{equation}
        \left(  
        \begin{array}{c}
                \hat{\eta}_8  \\
                \hat{\eta}_0
        \end{array}
        \right) =\Omega \left(
        \begin{array}{c} 
                \eta  \\
                \eta '
        \end{array}
        \right) ~,
        \qquad
        \Omega = \left(
        \begin{array}{cc}
                \cos\chi & \sin\chi  \\
                -\sin\chi & \cos\chi
        \end{array}    
        \right)~.
        \label{}
\end{equation} 
The relations between the $\eta_8$ and $\eta_0$
and physical fields $\eta,~ \eta'$ is then given by,
\begin{equation}
        \left(  
        \begin{array}{c}
                \eta_8  \\
                \eta_0
        \end{array}
        \right) = \Theta\left(
        \begin{array}{c} 
                \eta  \\
                \eta'
        \end{array}
        \right)~,
        \label{partr}
\end{equation}
where,
\begin{equation}
        \Theta = \Upsilon R \Omega = 
        \left(
        \begin{array}{cc}
    z\cos\lambda\cos\chi-f\sin\lambda\sin\chi &
      z\cos\lambda\sin\chi+f\sin\lambda\cos\chi  \\
         -z\sin\lambda\cos\chi-f\cos\lambda\sin\chi &
         -z\sin\lambda\sin\chi+f\cos\lambda\cos\chi
        \end{array}
        \right)~.
        \label{tetatr}
\end{equation} 
Note that the transformation $\Theta$ depends on two mixing angles 
$\lambda~\mbox{and}~\chi$ and two rescaling parameters $z~\mbox{and}~f$.
To write $\Theta$ in a more convenient and familiar form,  we define new
parameters,
\begin{eqnarray}
	z^2_1 = z^2 \cos^2\chi +f^2\sin^2\chi,~&~z\cos\chi = 
	z_1\cos\psi_1,~& ~f\sin\chi = - z_1\sin\psi_1,
	\label{¥} \\
	z^2_2 = f^2 \cos^2\chi +z^2\sin^2\chi,~&~f\cos\chi = 
        z_2\cos\psi_2,~ & ~z\sin\chi = - z_2\sin\psi_2,
	\label{¥} \\
	\tan\psi_1 = - \frac{f}{z}\tan\chi, ~ & ~	
        & ~
        \tan\psi_1 =   \frac{f^2}{z^2}\tan\psi_2,  
	\label{zpsi} \\                              
                       \theta_\eta = \lambda-\psi_2~,~ & ~
                       \qquad \theta_{\eta'} = \lambda-\psi_1~.
        \label{}  
\end{eqnarray}¥
and write $\Theta$ in the
form \ref{twoang}\cite{kaiser98,escrib99}
\begin{equation}
        \Theta  = 
    \left(\begin{array}{cc}
    \cos\theta_{\eta'} & ~\sin\theta_{\eta}  \\
         -\sin\theta_{\eta'} & ~\cos\theta_{\eta}
        \end{array}\right)\bar{Z}~,
        \label{tetatr1}
\end{equation}       
with $\bar{Z}=diag(z_1, z_2)$. This exact general form of the field 
transformation includes two
mixing angles $\theta_{\eta} ~\mbox{and}~ \theta_{\eta'}$, as well as two
rescaling parameters $z_1 ~\mbox{and}~z_2$.  A similar two mixing
angle scheme was proposed by Escribano and Fr\`{e}re \cite{escrib99}. 
In our scheme, however, the physical $\eta$ and $\eta '$ are
orthogonal.

We now turn to demonstrate that the pseudoscalar decay constants 
$F^i_P ~~(i=8,0;~P=\eta,~\eta ')$ are related to the octet and singlet
 decay constants $F_8,~\mbox{and}~F_0$, the same way the physical 
$\eta,~\eta '$ are related to the intrinsic $\eta_8,~\eta_0$ states. 
The axial vector currents read,
\begin{equation}
	\left(
	\begin{array}{c}
		J^8_\mu  \\
		J^0_\mu
	\end{array}
	\right) =  F K\left(
	\begin{array}{c}
		\partial_\mu \eta_8  \\
		\partial_\mu \eta_0
	\end{array}¥
	\right)~,¥
	\label{axcur}
\end{equation}¥
where $F=diag(F_8,~F_0)$. The matrix of the corresponding physical decay
constants  $F^i_{ph}$ is, 
\begin{equation}
	\left(
	\begin{array}{cc}
		F^8_\eta & F^8_{\eta'}  \\
		F^0_{\eta } & F^0_{\eta '}
	\end{array}¥
	\right) = F K \Theta = F \tilde{\Theta}~.
	\label{¥}
\end{equation}
Since,
\begin{equation}
K \Upsilon R = \Upsilon R^{-1}~,
\end{equation}
one obtains,
\begin{equation}
	\tilde{\Theta} = \Upsilon R^{-1} \Omega = \sqrt{\kappa_8\kappa_0}
	\left(
        \begin{array}{cc}
    f\cos\lambda\cos\chi-z\sin\lambda\sin\chi &
      f\cos\lambda\sin\chi+z\sin\lambda\cos\chi  \\
         -f\sin\lambda\cos\chi-z\cos\lambda\sin\chi &
         -f\sin\lambda\sin\chi+z\cos\lambda\cos\chi
        \end{array}
        \right)~.
	\label{¥}
\end{equation}¥
Again we introduce new parameters, 
\begin{eqnarray}
	\tilde{z}^2_1 = f^2 \cos^2\chi +z^2\sin^2\chi,~&~f\cos\chi = 
	\tilde{z}_1\cos\tilde{\psi}_1,~& ~z\sin\chi = - 
	\tilde{z}_1\sin\tilde{\psi}_1,
	\label{¥} \\
	\tilde{z}^2_2 = z^2 \cos^2\chi +f^2\sin^2\chi,~&~z\cos\chi = 
   \tilde{z}_2\cos\tilde{\psi}_2,~ & ~f\sin\chi = - 
   \tilde{z}_2\sin\tilde{\psi}_2,
	\label{¥} \\
	\tan\tilde{\psi}_1 = - \frac{z}{f}\tan\chi,~ & ~	
        & ~
        \tan\tilde{\psi}_2 = \frac{f^2}{z^2}\tan\tilde{\psi}_1.
	\label{tilzpsi}
\end{eqnarray}
From Eqns. \ref{tilzpsi} and \ref{zpsi}  
$\tilde{\psi}_1 = \psi_2~\mbox{and}~\tilde{\psi}_2 = \psi_1$   
and after simple algebraic manipulations one obtains,
\begin{equation}
       \tilde{\Theta} =  \Lambda\Xi,~~~~\Lambda =  
      \left(\begin{array}{cc}
      \cos\theta_{\eta} & ~\sin\theta_{\eta'}  \\
      -\sin\theta_{\eta} & ~\cos\theta_{\eta'}
        \end{array}\right) ~,
        \label{mlambda}
\end{equation} 
with $\Xi = diag(\xi_1, \xi_2) = diag(\tilde{z}_1 \sqrt{\kappa_8
\kappa_0}, ~\tilde{z}_2\sqrt{\kappa_8 \kappa_0})$. Hence the 
pseudoscalar decay constants mix according to,
\begin{equation}
	\left(
	\begin{array}{cc}
		F^8_\eta & F^8_{\eta'}  \\
		F^0_{\eta} & F^0_{\eta '}
	\end{array}
	\right)  = {F} \Lambda\Xi~,
	\label{decconst}
\end{equation}
 This relation should be compared with the corresponding relation 
 for the fields (see Eqns.\ref{partr},\ref{tetatr1}),
\begin{equation}
        (\begin{array}{cc}
                \eta & \eta'
        \end{array}) = (\begin{array}{cc}
        \tilde{\eta}_8 &\tilde{\eta}_0
        \end{array})\Lambda \bar{Z}^{-1}~,
        \label{}
\end{equation}
where,
\begin{equation}
        (\begin{array}{cc}
        \tilde{\eta}_8 & \tilde{\eta}_0
        \end{array}) = (\begin{array}{cc}
        \eta_8/ \Delta  & \eta_0/ \Delta 
        \end{array})~,
        \label{}   
\end{equation}     
and $\Delta =
\cos(\theta_{\eta'}-\theta_{\eta})$.
 We may thus conclude that the $renormalized$ decay constants and
fields follow the same transformation $\Lambda$.  Clearly, even in the
limit of exact nonet symmetry, where $F_8 = F_0$, the reduction of the
meson kinetic term to the standard quadratic form may lead to
renormalized decay constants $\tilde{F}_8 \neq \tilde{F}_0$.

It must be stressed again that, for any EFT for which the kinetic term
has a bi-linear form, the mixing schemes for the states and decay constants
involve two angles and two rescaling parameters. 
Clearly, for a diagonal kinetic matrix, $\lambda = 0$, 
$\psi_1$ and  $ \psi_2 $ vanish, and the results reduce to a one 
angle mixing scheme (Eqn.\ref{oneang}). The scheme of Feldmann et
al.\cite{feld98,feld98a,feld98b} corresponds to ideal mixing where
flavorless mesons are represented by fields with or without strange
quark content. Their assumption that, the states and decay constants
follow the same one angle mixing scheme (Eqns.\ref{ffeld}-\ref{mfeld})
is equivalent to the assumption that the kinetic term has the standard 
form. This immediately leads to one mixing angle scheme
with neither field nor decay constant renormalizations.
 
In  summary for any EFT the particle and decay constant mixing is 
defined solely by the effective kinetic and mass terms of the particle
 Lagrangian. Since radiative corrections may only alter 
the values of the matrix elements of $K$ and $M^2$ but do not change 
the general form of $L_{km}$ this statement is valid for any 
type of interactions the theory accounting for. Previously suggested
mixing schemes 
correspond to a proper choice of the kinetic matrix. 
Particularly, $\eta-\eta '$ mixing require one mixing angle
scheme, if and only if, the kinetic term for the intrinsic fields 
has a quadratic form. Such are the traditional scheme \cite{pdg00} 
and the one proposed by Feldmann et al.\cite{feld98,feld98a,feld98b}. 
A bi-linear (i.e. with non diagonal matrix) kinetic term leads to a 
two mixing angle scheme similar but not identical to the one proposed by
Escribano and Fr\`{e}re\cite{escrib99}.

\bigskip
{\bf Acknowledgment}  This work was supported in part by the Israel
Ministry of Absorption.

\end{document}